\documentclass[10pt,aps,pra,amssymb,a4paper,preprint,twocolumn,tigthen,footinbib]{revtex4}
\usepackage{latexsym,amssymb,amsfonts,amsbsy}
\def\vec#1{\boldsymbol{#1}}
\newif\ifpdf
\ifx\pdfoutput\undefined
\pdffalse 
\else
\pdfoutput=1 
\pdftrue
\fi
\ifpdf
\usepackage[pdftex]{graphicx}
\DeclareGraphicsExtensions{.pdf}
\else
\usepackage[dvips]{graphicx}
\DeclareGraphicsExtensions{.eps}
\fi
\usepackage{times}
\newcommand{\p}{\ensuremath{\mathrm{p}}}
\newcommand{\e}{\ensuremath{\mathrm{e}}}
\renewcommand{\d}{\ensuremath{\mathrm{d}}}
\renewcommand{\H}{\ensuremath{\mathrm{H}}}
\newcommand{\Ps}{\ensuremath{\mathrm{Ps}}}
\newcommand{\He}{\ensuremath{\mathrm{He}}}
\catcode`\@=11
\def\eqalign#1{\null\,\vcenter{\openup\jot\m@th
\ialign{\strut\hfil$\displaystyle{##}$&$\displaystyle{{}##}$\hfil
     \crcr#1\crcr}}\,}
\catcode`\@=12
\begin{document}
\title{Borromean binding}
\thanks{Dedicated to my colleague and friend Vladimir Belyaev 
at the occasion of his 70$^\mathrm{th}$ birthday}%

%
\preprint{\begin{minipage}{4cm}LPSC-03-09\\
arXiv/nucl-th/0305076\end{minipage}}
\author{Jean-Marc Richard}
\address{Laboratoire de Physique Subatomique et Cosmologie\\
 Universit\'e Joseph Fourier--CNRS-IN2P3\\
53, avenue des Martyrs, 38026 Grenoble, France}

\begin{abstract}\noindent
A review is first presented of the Hall--Post inequalities relating 
$N$-body to $(N-1)$-body energies of quantum bound states. These inequalities
 are then applied  to delimit, in the space of coupling constants, the domain of Borromean 
 binding where a composite system is bound while smaller subsystems are unbound.
\end{abstract}
\maketitle
\section{Introduction}\label{se:Intro}
%
There are many examples, at various scales, 
of composite systems at the edge between binding and non-binding. 
In nuclear physics, a proton--proton or neutron--neutron pair misses binding by 
 a small margin, while a proton and
a neutron form a rather weakly bound deuteron. 
The existence of a near-threshold state can induce dramatic consequences, for instance on fusion
probabilities \cite{Motovilov:1999ef}.
A pair of charmed mesons is presumably near the border separating
stability from spontaneous dissociation \cite{Ericson:1993wy}. Atoms such as ${}^4\He$ were 
for a long time believed to be unable to merge into a
molecule. 
Recent studies indicates a tiny binding of the order of $1\;\mathrm{mK}$ 
for ${}^4\He_{2}$. 
However, if one replaces one of the ${}^4\He$
 by an atom containing the lighter isotope ${}^3\He$, 
 then the ${}^3\He{}^4\He$ is unbound. 
 For a recent review on  ${}^3\He_{N}{}^4\He_{M}$ systems, 
 see, e.g., Refs.~\cite{GuaPRL,BressPRL}.

An intriguing  question is whether it is easier to bind three or
more  components  than to form a mere two-body bound state. 
An answer is provided by the study of halo nuclei, 
which contain peripheral neutrons. 
Consider for instance the ${}^6\He$ nucleus. 
It is stable against  any dissociation, while the lighter ${}^5$He 
spontaneously decays into a neutron and a ${}^4$He. 
In the (reasonable) approximation where the structure of the core 
is neglected, this means that the $(\alpha,\,\mathrm{n},\,\mathrm{n})$ 
three-body system is bound, while neither 
$(\alpha,\,\mathrm{n})$ nor $(\mathrm{n},\,\mathrm{n})$  have a discrete spectrum.

\begin{figure}[b]
\begin{center}
\includegraphics[width=.2\textwidth]{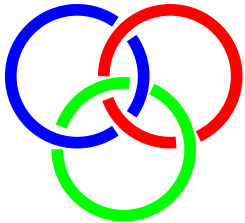}
\end{center}\caption{\label{BorroFig} Borromean rings}
\end{figure}

This property of 3-body binding without 2-body binding was astutely
named {\em Borromean} \cite{Bang93}, 
after the Borromean rings, which are interlaced in  a subtle 
topological way (see Fig.~\ref{BorroFig}) such that if any one of them is removed, 
the two other become unlocked. 
The adjective {\em Borromean} is nowadays broadly accepted 
in the field of quantum few-body systems.

Borromean binding is intimately related to two other
fascinating properties of few-body quantum systems. The Efimov
effect \cite{Efimov70} indicates that when the two-body energy 
vanishes (e.g., by tuning the strength of the potential),
a myriad of weakly-bound states show up in the three-body spectrum. 
This implies that the three-body ground-state already exists
at this point. Slightly above the onset of two-body binding, 
the ratio $E_{2}/E_{3}$ of two-body to three-body
binding energies is very small. By  rescaling, one can
reach a situation with a finite 2-body energy, 
and a 3-body energy that becomes infinite when the range of the potential 
is made shorter and shorter: this is the
Thomas collapse \cite{Thomas35}.

This review is organised as follows. 
In Sec.~\ref{se:HP}, the Hall-Post inequalities are briefly
recalled. They are applied in Sec.~\ref{se:BorroB} to constraint the
domain of coupling constants leading to
Borromean binding for bosons interacting through
short-range forces.  The difficulties arising in the case of fermions are described in  
Sec.~\ref{se:HPF}. Borromean
binding with Coulomb forces is the subject of Sec.~\ref{se:BorroC}, before the conclusions.
\section{Hall--Post inequalities}\label{se:HP}
A number of inequalities can be written down for binding energies in 
quantum mechanics if one splits the Hamiltonian
into pieces (each piece being hermitian). Thus, for example,
\begin{equation}\label{basic-split}
H=A+B+\cdots\ \Rightarrow\  E(H)\ge E(A)+E(B)+\cdots~,
\end{equation}
in an obvious notation where $E(H)$ is the ground-state energy of $H$. 
Saturation is obtained if $A$, $B$, etc., reach
their minimum simultaneously.
If, for instance, $H=\vec{p}^2-1/r+r^2/2$
describes the motion of a particle feeling both a Coulomb and an harmonic potential,
then $E(H)\ge (-1/2) + (3/2)$, corresponding to an equal share of the kinetic energy.
A slight improvement is obtained by writing
$H=\left[\alpha \vec{p}^2 -1/r\right]+\left[(1-\alpha)\vec{p}^2+r^2/2\right]$, 
and optimising $\alpha$.

The reasoning can be applied to obtain a lower bound on 
3-body energies in terms of 2-body energies. 
This has been discovered independently by several authors 
working on the stability of
matter \cite{FIRU66} or baryon spectroscopy in simple 
quark models \cite{Nussinov:1999sx}. Let
\begin{equation}
\label{three-body}
H_N(m,g)=\sum_{i=1}^N {\vec{p}_i^2\over 2 m}+g \sum_{i<j} V(r_{ij})~,
\end{equation}
be the Hamiltonian describing a system of $N$ identical particles interacting 
with pairwise forces, and  $E_{N}(m,g)$ be its ground state energy .
From the identity
\begin{equation}
\label{decom-1}
H_N(m,g)=\sum_{i}H_{N-1}^{(i)}\left((N-1)m,{g\over N-2}\right)~,
\end{equation}
where $H_{N-1}^{(i)}$ does not include particle $i$, one derives 
\begin{equation}\label{HP1}\eqalign{
E_N(m,v)&\ge N E_{N-1}\left((N-1)m,{g\over N-2}\right)\cr
&\ge {N(N-1)\over2}E_{2}(m(N-1),g))~,}
\end{equation}
which can supplement any upper bound provided by a variational method, to frame the exact energy. 
The generalisation to unequal masses or different potentials among the pairs 
is straightforward. For instance,
\begin{equation}
H_{3}(\{m_{i}\};\{g_{ij}\})=
\left[{\vec{p}_{1}^2\over 4m_{1}}+{\vec{p}_{2}^2\over 4m_{2}}+g_{12}V(r_{12})\right]+\cdots
\end{equation}

Numerical investigations show that the lower bound (\ref{HP1}) is not 
very accurate and never reaches saturation. The relative wave function of particles 1 and 2 
is of course modified by the presence of the third particle, 
and thus the three-body wave function does not describe
optimally the (1,2) subsystem. However, this is a small effect which even 
disappears for the harmonic oscillator.
The main source of inaccuracy in (\ref{HP1})
 is that the contribution of the (1,2) subsystem is replaced by the {\em rest} 
energy of an isolated (1,2) pair, although  the
(1,2) pair is  not at rest in the whole system.

To overcome this difficulty, a better decomposition
was written down by Hall and
Post \cite{HP67}, and rediscovered in Ref.~\cite{BMR12}. It involves the 
translation-invariant part $\widetilde{H}_{N}$
\begin{equation}
\label{Def-tilde }
\widetilde{H}_N=H_{N}-{(\vec{p}_1+\vec{p}_2+\cdots)^2\over 2Nm}~,
\end{equation}
of each Hamiltonian $H_{N}$ and reads
\begin{equation}
\label{decom-2}
\widetilde{H}_N(m,g)=\sum_{i<j}\widetilde{H}_{N-1}^{(i)}
\left(m{N(N-2)\over N-1},{g\over N-2}\right)~.
\end{equation}
This leads to the new inequality
\begin{equation}\label{HP2}\eqalign{
E_{N}(m,g)&\ge E_{N-1}\left(m{N(N-2)\over N-1},{g\over N-2}\right) \cr
&\ge{N(N-1)\over2}E_{2}\left({mN\over2},g\right)~,}
\end{equation}
which is necessarily better than (\ref{HP1}) since, $mN/2<m(N-1)$ for $N\ge3$, and 
for a given potential, the
 binding energy is a decreasing function of the constituent mass.
This inequality is optimal in the sense that it can be saturated: this occurs for the harmonic oscillator and only in
this case.
The inequality (\ref{HP2}) also holds for the 
variational approximation $E_N^\mathrm{var}$ to the binding energy $E_N$, 
provided the sets of trial functions $\psi$ for $N=2$ and $N=3$ 
are consistent. 
In particular, one gets saturation, 
$E_3^\mathrm{var}(m,v)=3 E_2^\mathrm{var}(3m/2,v)$ if each 
variational energy is calculated with a single harmonic-oscillator function 
$\psi_N\propto\exp(-\alpha_N\sum r_{ij}^2)$.
This is a rather frequent pattern that (with minimal restrictions) 
variational solutions share the properties of
the exact ones. For the virial theorem this was underlined   
by Fock and rediscovered by many authors \cite{Fock}.

The case of particles with unequal masses has revealed some surprises. 
The simple extension (here for $N=3$) of 
 the decomposition (\ref{decom-2})
\begin{equation}\label{decom-3}
\widetilde{H}_3[m_i;V_{ij},\ldots]=
\alpha_{12}\left({m_1 \vec{p}_2-m_2 \vec{p}_1\over m_1+m_2}\right)^2+V_{12} +
\cdots~,
\end{equation}
does not always give saturation for the harmonic oscillator. 
The remedy was found in Ref.~\cite{BMR3} for $N=3$ particles and extended 
in Ref.~\cite{Benslama98} for $N=4$: one should introduce more freedom 
in the decomposition of the Hamiltonian and find the optimal choice 
by suitable adjustment.

For $N=3$, the decomposition is written as
\begin{equation}\label{decom-4}
H_{3}=\vec{B}.\vec{P}+
\alpha_{12}\left({\vec{p}_{1}-x_{3}\vec{p}_{2}\over 1+x_{3}}\right) + 
g V_{12}+\cdots~,
\end{equation}
where $\vec{P}=\sum\vec{p}_{i}$.
The identification gives the irrelevant vector $\vec{B}$ and the inverse masses 
$\alpha_{ij}$ in terms of the free parameters $x_{i}$. The lower bound
on $E_{3}$ is expressed as a sum of 2-body energies.  It depends on these $x_{i}$ 
and is maximised by varying these
parameters, for each given set of potentials $V_{ij}$. In the case of the harmonic oscillator with unequal masses and
even unequal strength factors, saturation is reached, i.e., the exact energy always coincides with the lower bound.

For $N=4$, one should go a step further, and write the decomposition as 
\begin{equation}\label{decom-5}
H_{4}=\vec{B}.\vec{P}+\alpha_{12} 
\left(\sum_{i} x_{12,i}\,\vec{p}_{i}\right)^2 + g V_{12}+\cdots
\end{equation}
The momentum $\sum_{i} x_{12,i}\,\vec{p}_{i}$ is constrained by translation invariance, and
$x_{12,2}-x_{12,1}=1$, to make it conjugate to $\vec{r}_{2}-\vec{r}_{1}$. 
This leaves two free parameters for each
pair, which are adjusted to optimise the lower bound.

Refs.~\cite{BMR12,BMR3,Benslama98} contain several examples, where the three- or four-body energy is computed and
compared to the above lower bounds.
For confining potentials, one is not too far from the 
limiting case of the harmonic oscillator for which the inequality 
is saturated, and the inequalities (\ref{HP2}) provide
a very good approximation to the exact energy. 
The quality of the approximation deteriorates for short-range or Coulomb forces.
\section{Borromean binding of bosons}\label{se:BorroB}
The Hall--Post decomposition was used in the previous 
section to obtain inequalities on energies at 
given coupling constant $g$. They can also be used to 
provide relations between  coupling constants at given energy $E$.
In particular for $E=0$, one can derive inequalities on the critical couplings $g_{N}$, 
where $g_{N}$ is the minimal strength required to achieve $N$-body binding.

In one- or two-dimensional quantum mechanics, any attractive potential 
supports at least one bound state \cite{bsdim12}.
More precisely, $g_{2}=0$ if the integral $\int \d^n x V(\vec{x})$ is negative.
For $n=3$ dimensions, the problem is more subtle \cite{Landau77}. An attractive potential
with short range, say $g V(r)$, requires a minimal strength $g>g_{2}$ 
to achieve binding of two constituents of unit mass, and 
$mg >g_{2}$ for constituents with mass $m$. A classic paper by
 Blatt and Jackson \cite{Blatt49} gives estimates of $g_{2}$ for simple potentials such as Yukawa.
 
In this context, the phenomenon of Borromean binding is expressed by the property that the 
critical coupling $g_{3}$ required to bind three bosons is {\em smaller} than $g_{2}$. More generally,
$g_{N}\le g_{N-1}$. Several questions can be raised: what is the typical range of values for $g_{3}/g_{2}$, 
$g_{4}/g_{3}$, etc.? Are there rigorous constraints on these ratios? Are these constraints saturated
for particular potentials? What can be said in the limit of large $N$?

Some results are given in Refs.~\cite{Richard94,Goy95,Mosz00}, which contain references to earlier papers.
In particular, the decomposition (\ref{decom-2}) 
$H_{N}\propto\sum H_{N-1}$ implies that $H_{N}$ hardly explores the domain 
of binding with negative expectation values, if all $H_{N-1}$ remain positive. Thus
\begin{equation}\label{Boroo1}
N g_{N}\ge (N-1)g_{N-1}\ge \ldots 2 g_{2}~.
\end{equation}
For simple monotonic potentials, one  gets $g_{3}/g_{2}\simeq0.80$, 
well above the rigorous bound $g_{3}/g_{2}\ge 2/3$.
Still, there is a remarkable window of about $20\%$  in the coupling constant, 
where three-body systems are bound without two-body
binding. One also gets $g_{4}/g_{2}\simeq 0.67$, typically, which, when compared 
to $g_{3}/g_{2}\simeq0.80$, reveals a window 
of about $13\%$ for four-body binding without
three-body binding.

For bosons, the situation is rather simple. If a potential 
does not succeed in binding a pair of bosons, but is
predominantly attractive, that is to say has a 
negative scattering length,  it will bind a sufficiently large number of bosons.

We have seen that Hall--Post inequalities are saturated for harmonic potentials. 
A pure $V(r)\propto r^2$ does not give Borromean binding, 
as $g V$ would confine even for $g\to 0+$. But a potential 
like $V\propto r^2- \mathrm{C}$ in the domain of the relevant wave functions, and vanishing at very large distances $r$,
would give $g_{3}/g_{2}\simeq 2/3$, i.e., nearly saturate the bound~\cite{Mosz00}. 
This corresponds to an attractive potential with an external barrier.

More important for physical applications are potentials with an internal repulsive core.
Their wave functions 
significantly depart from those of the harmonic oscillator
and the ratio $g_{3}/g_{2}$ become closer to 1. Examples are given in \cite{Mosz00}.
\section{The case of fermions}\label{se:HPF}
In general, an inequality such as (\ref{HP2}), though remaining valid, is of little interest for
fermions. The ground state $E_{2}$ corresponds to a symmetric orbital wave function, 
with, for instance, the spins arranged in a singlet state to fulfil  the antisymmetry requirement. On the other hand, 
three or more electrons can never be in a fully-symmetric  orbital wave function. Thus the actual $E_{N}$, $N\ge3$, 
will be an excited level of the $E_{N}$ entering the inequality (\ref{HP2}).

A minimal knowledge of the structure of the wave function is thus needed to  write down an useful
inequality. Consider for instance three spin 1/2 fermions in a symmetric spin 
(or spin--isospin, or spin--isospin--colour) state.
Then the space wave function of the ground state should be a $L^P=1^+$ 
antisymmetric wave function, a prototype  being the
harmonic oscillator state 
$\vec{\rho}\times\vec{\lambda}\exp[-\alpha(\rho^2+\lambda^2)]$ in terms of 
 Jacobi coordinates
$\vec{\rho}$ and $\vec{\lambda}$. The two-body subsystems 
are in an antisymmetric state with orbital momentum
$\ell=1$. Then $E_{3}(m,g)\ge 3 E_{2}(3m/2, g)$ results from  the identity
$H_{3}(m,g)=\sum H_{2}(3m/2, g)$ applied in the subspace of 
antisymmetric wave functions.

For more than $N=3$ fermions, or  other spin states of $N=3$, 
the problem is more complicated 
but still, some rigorous results can be obtained. 

The basis is the decomposition of $N$-particle representations of the
permutation group in terms of $N-1$ states \cite{Juillet:2000ie}. 
For instance, in a spin $S=1/2$ state of three
electrons a given pair is half of the time in a singlet state, and half of time in a triplet state. This means
a proper bound on $E_{3}$ involves, for $E_{2}$, and average over the $\ell=0$ and the $\ell=1$ ground state.
For other cases, one needs some group theoretical calculations \cite{Juillet:2000ie}.
This gives nice {\em local} constraints, i.e.,
links from $E_{N}$ to $E_{N-1}$ 
 (see also the improvements proposed in Ref.~\cite{VanNeck:2001bw}). 
But this approach fails when extrapolated 
from $E_{2}$ to the large-$N$ limit of $E_{N}$.
This means that much progress remains to be made. 
Checking the constraints with exactly solvable potentials
\cite{Khare01} gives valuable insight into the problem.
 
 Note that the question of writing inequalities for fermion energies  is rather old. In the course of the pioneering 
papers on the stability of matter, a very clever decomposition was written down
\begin{equation}
\eqalign{%
H_{N+1}=&\sum_{i} h_{N}^{(i)}~,\cr
h_{N}^{(i)}=&
\sum_{\alpha}{\mathstrut}^{\!(i)}{\vec{p}^2_{\alpha}\over2Nm}
+{g\over N-1}\sum_{\alpha<\beta}{\mathstrut}^{\!(i)}V_{\alpha\beta}~,}
\end{equation}
where the superscript in $\sum{}^{\!(i)}$ indicates that particle
 $i$ is omitted. Each $h_{N}^{(i)}$ is an independent-particle Hamiltonian
with ground-state energy $\sum_{j}\epsilon_{j}$, in terms of one-particle energies $\epsilon_{j}$, properly piled up
as  explained in textbooks of elementary chemistry (2 particles in 1s, 6 in 2p, etc.), to ensure antisymmetry. However, the recoil
of, e.g., the $\{2,\,3\,\ldots,\,N\}$ subsystem against the first particle is not taken into account. As a consequence,
the inequality is never saturated.

Significant progress was achieved by 
Basdevant and Martin \cite{Basdevant:1996mc}, who used sophisticated convexity
inequalities to derive a bound for power-law potentials, 
that becomes saturated in the limit of harmonic forces.

\section{Borromean molecules}\label{se:BorroC}
Borromean binding as described in Sec.~\ref{se:BorroB} seems at first hardly conceivable for a Coulomb potential, since
changing the
strength by an overall factor $g$ simply results in a mere rescaling by  $g^2$ of all binding energies
$E_{N}$.
However, if the stability of ions and molecules does not depend on the overall strength of the interaction, it relies on
an adequate balance between  attraction and repulsion, and,  for a given set of charges, on the ratios of the
constituent masses which are involved. 

Consider for instance a set  of three masses $\{m_{i}\}$ carrying 
 charges $\{q_{i}\}=\pm\{+1,-1,-1\}$, (times  an overall
factor).
Systems with identical or nearly identical inverse masses $m_{2}^{-1}$ and $m_{3}^{-1}$, such as $\H_{2}^+\,
(\e^-,\p,\p)$, $\Ps^-\, (\e^+,\e^-,\e^-)$ or $\H^-\, (\p,\e^-,\e^-)$, 
or neighbouring configurations, are stable, while less symmetric systems such as $(\e^-,\p,\e^+)$ or
$(\p,\bar\p,\e^-)$ spontaneously decay into an atom and an isolated charge \cite{ThreeBodyCoul}.

Note that the stable systems with three  unit charges are not Borromean, since
 there are always two stable atoms among the subsystems, $\{1,2\}$ and 
$\{1,3\}$ in our notation. This
means that one can construct each stable system step by step: two charges form a stable atom, 
which attracts the third charge.

For $N=4$ unit charges $\{q_{i}\}={+1,+1,-1,-1}$, there are new possibilities. 
The most familiar cases are the following:
\begin{itemize}\itemsep-2pt
\item
unstable configurations, such as $(\p,\e^+,\bar{\p},\e^-)$\footnote{In this section,we consider Coulomb forces only,
 and annihilation, strong interaction, etc., are neglected}, which spontaneously splits into a protonium and a
positronium,
\item
stable systems, the simplest being  the $\Ps_{2}$ molecule $(\e^+,\e^+,\e^-,\e^-)$, whose  
all three-body subsystems $\pm(\e^+,\e^-,\e^-)$  are stable,
\item
less stable systems, such as the positronium hydride $(\p,\e^+,\e^-,\e^-)$, which can be viewed as a stable
$\H^-$ fixing a positron, or a stable $\Ps^-$ attached to a proton, or, more interestingly, as an unstable $(\e^-
,\p,\e^+)$ stabilised by the addition of a second electron.
\end{itemize}
More recently, a new category of four unit-charge systems was revealed \cite{Richard03}. They are stable, but 
{\em all} three-
body subsystems are unstable. Hence, they are Borromean, in the sense that they cannot be built by adding the
constituents one by one. An example is $(\p,\d,\bar{\p},\bar{\d})$, with a proton, a deuteron and their antiparticles.
Indeed, a study by Mitroy \cite{Mitroy2000}, who uses Varga's stochastic variational method \cite{Varga95}, 
indicates that asymmetric ions $(m^+,M^-,m^-)$ are stable in the range 
\begin{equation}\label{Mitroy}
0.70\lesssim {M\over m} \lesssim1.69~,
\end{equation}
while other studies, by Bressanini et al.\ \cite{Bressanini97}, who use a diffusion Monte-Carlo approach, 
or Varga \cite{Varga99} with a stochastic search of the range coefficients of a Gaussian parametrisation, 
have established that $(M^+,m^+,M^-,m^-)$ remain stable for 
\begin{equation}\label{MmMm}
{1\over 2.2}\lesssim {M\over m}\lesssim2.2~.
\end{equation}
Then, for $M/m\sim 2$, and for neighbouring configurations $(M^+,m^+,M'^-,m'^-)$, 
the four-body system is stable
while none of its three-body subsystems is stable.

\section{Outlook}
In molecular, nuclear, or hadronic physics, 
there are fragile and subtle quantum states, whose stability relies on the
joint efforts of many subsystems, which by themselves do not contain enough 
attraction to be stable. It is of
course tempting to extend to political or sociological situations: \textsl{l'union fait la force}, 
as it is said in a popular French proverb.

\begin{acknowledgments} 
I would like to thank my collaborators on the subjects evoked here: 
Jean-Louis Basdevant,
Sonia Fleck, J\'er\^ome Goy, Olivier Juillet, Avinash Khare, 
Ali Krikeb, Andr\'e Martin, Kalman Varga, Steve Moszkowski,
Tai T.~Wu, as well as Sami Zouzou and his colleagues of Constantine. 
I also thanks A.J.~Cole for his comments on the manuscript.
Pr.~Motovilov is gratefully acknowledged for inviting 
me to contribute to this book. 
It is a great pleasure to remember long and fruitful discussions 
with Vladimir Belyaev, in particular at Grenoble and  Trento: his enthusiam is catching
 and his approach to physics problems elegant and penetrating.
\end{acknowledgments} 

\end{document}